\documentstyle[12pt]{article}

\def\be{\begin{equation}}
\def\ee{\end{equation}}
\begin{document}
\setcounter{page}{1}
\pagestyle{plain}

\vspace{1cm}
\begin{center}
\Large{\bf Bandgap Narrowing in Quantum Wires  }\\
\small \vspace{2cm}   {\bf Kourosh  Nozari}\quad and\quad {\bf Mahyar Madadi}\\
{\it Department of Physics,
Faculty of Basic Sciences,\\
University of Mazandaran,\\
P. O. Box 47416-1467,
Babolsar, Iran\\
e-mail: knozari@umz.ac.ir}\\
\end{center}
\vspace{1.5cm}

\begin{abstract}
In this paper we consider two different geometry of quasi
one-dimensional semiconductors and calculate their
exchange-correlation induced bandgap renormalization (BGR) as a
function of the electron-hole plasma density and quantum wire width.
Based on different fabrication scheme, we define suitable external
confinement potential and then leading-order GW dynamical screening
approximation is used in the calculation by treating
electron-electron Coulomb interaction and electron-optical phonon
interaction. Using a numerical scheme, screened Coulomb potential,
probability of different states, profile of charge density and the
values of the renormalized gap energy are calculated and the effects
of variation of confinement potential width and temperature
are studied.\\
{\bf PACS}: 73.20.Dx; 71.35.Ee; 71.45.Gm\\
{\bf Key Words}: Low Dimensional Structures, Quantum Wires, Bandgap
Renormalization, V-Shaped and T-Shaped Confinement Potentials\\
\end{abstract}
\newpage

\section{Introduction}
A highly dense electron-hole plasma can be generated in a wide
variety of semiconductors by optical pumping. The band structure and
the optical properties of highly excited semiconductors differ from
those calculated for non-interacting electron-hole pairs due to
many-body exchange-correlation effects arising from the
electron-hole plasma [1-3]. One of the important many-body effect in
high density electron-hole plasma is a density-dependent
renormalization of the fundamental band gap of the semiconductor,
which causes an increasing absorption in the spectral region below
the lowest exciton resonance. The exchange-correlation correction of
the fundamental band gap due to the presence of free carriers
(electrons in the conduction band and holes in the valence band) in
the system is referred to as the band gap renormalization (BGR)
effect. Optical nonlinearities, which are strongly influenced by
Coulomb interaction in the electron-hole plasma, are typically
associated with the bandgap renormalization phenomenon.\\
The band gap re-normalization has been widely studied in bulk and
quasi-two dimensional (quantum well) semiconductors [4-6]. In recent
years, quasi-one dimensional semiconductor quantum wires (QW) have
been fabricated in variety of geometric shape with atomic scale
definition, and QW optical properties have been studied for their
potential device applications such as semiconductor lasers[7-9].
There has, however, been little work on the BGR in QW systems, both
experimentally and theoretically. Recently different geometries of
quantum wires, such as rectangular, V-shaped and T-shaped quantum
wires have been fabricated and studied, and various experimental
technics for fabrication and growth of these structures have been
developed[2,3]. Square quantum well wires have been studied by Hu
and Das Sarma[10]. They have calculated the value of the band gap
re-normalization for this case in GW approximation. Band gap
renormalization in photoexcited semiconductor quantum wire in GW
approximation has been studied by Hwang and Das Sarma[11]. Rinaldi
and Cingolani have studied optical properties of 1D quantum
structures specially the case of V-shaped quantum wire[12]. They
have considered confinement potential of the form $V(y) =
-\frac{V_{0}}{\cosh^{2}(\alpha y)}$ but they have not calculated the
band gap re-normalization with this confinement potential
theoretically. Bener and Haug have considered plasma-density
dependence of the optical spectra for quasi-one-dimensional quantum
well wires[13]. Tanatar has studied band gap re-normalization in
quasi-one dimensional systems in simple plasmon-pole (quasi-static)
approximation[14]. On the other hand, T-shaped quantum wires
recently have been considered by some authors. For example Sedlmaier
and his coworkers have studied band gap re-normalization of
modulation doped T-shaped quantum wires. They have presented
self-consistent electronic structure calculations for this device.
These calculations show a band gap re-normalization which, when
corrected for excitonic energy and its screening, are largely
insensitive to the 1D electronic density[15]. Stopa, using density
functional theory, has calculated the electronic structure of a
modulation doped and gated T-shaped quantum wire. He also has
calculated the band gap re-normalization as a function of the
density of conduction band electrons[16]. Lin, Chen and Chuu have
found the dependence of the bound states of L-shaped and T-shaped
quantum wires to some asymmetric parameter in an inhomogeneous
magnetic fields[17]. Sedlmaier and his coworkers have calculated the
band-gap re-normalization of modulation doped quantum wires by
considering the photoluminescence spectra as a function of the one
dimensional density specially in T-shaped quantum wires[18]. Nozari
and Madadi, recently have proposed a theoretical framework for
calculation of band gap re-normalization in V-shaped quantum wires.
They have calculated numerically the value of this re-normalization
in various temperatures and carrier density and have found exact
solutions for some geometry of this type of quantum wires in a
simplified random phase approximation[19]. So far most of
calculations have been done in the static screening approximation or
in the simple plasmon-pole approximation, which is a simplified
version of the random-phase approximation (RPA). The plasmon-pole
approximation consists of ignoring the weight in the single particle
excitations and assuming that all free carrier contributions to the
dynamical dielectric function lies at the effective plasma
frequency. The advantages of the plasmon-pole approximation are its
mathematical simplicity and simple physical meaning. However, a
certain degree of arbitrariness in the choice of the effective
plasmon pole parameters leads to considerable difficulties in
applying the theory to semiconductors with complex band structures.
In this paper, we calculate the BGR of the V-shaped and T-shaped
quantum wire structures based on the RPA dynamical screening (GW)
scheme by taking into account special mathematical definitions for
confinement potentials which differ considerably from existing
literature specially [12] and [19]. In doing so, the full frequency
dependent dielectric response in the two component one-dimensional
electron-hole plasma is considered.\\
The structure of the paper is as follows: In section 2 we provide
formal theory of bandgap renormalization in a general quasi
one-dimensional structure using dynamical random phase
approximation. Section 3 considers V-shaped confinement potential.
First a suitable mathematical definition for this form of
confinement potential is given. Then a numerical scheme for
solutions of exact equations is provided and various quantities are
calculated numerically. Section 4 considers the same analysis for
T-shaped confinement potential. Conclusions and discussion are given
in section 5.

\section{BGR in Dynamical Random Phase Approximation}
The exchange-correlation induced correction of the fundamental band
gap due to the presence of free carriers (electrons in the
conduction band and holes in the valence band) in the system is
referred to as the band gap re-normalization effect. The BGR is
given by the sum of the self-energies for electrons and holes at
band edges
\begin{equation}
\label{math:2.1} \Delta = Re\Big(\Sigma_{e}(0, 0)\Big) +
Re\Big(\Sigma_{h}(0, 0)\Big),
\end{equation}
where $Re$ stands for real part. The total electronic self-energy
within the leading order effective dynamical interaction(GW) in a
two-component electron-hole plasma is
\begin{equation}
\label{math:2.2} \Sigma_{e}(k,\omega) =
i\int\frac{dq}{2\pi}\int\frac{d\omega^{\prime}}{2\pi}G_{0}(k-q,
\omega-\omega^{\prime})\frac{V_{s}(q,\omega^{\prime})}{\epsilon(q,\omega^{\prime})},
\end{equation}
where $G_0$ is the Green's function for noninteracting electron gas,
$V_{s} = V_{c} + V_{ph}$ the total effective interaction and
$\epsilon(q,\omega) = 1-V_{s}(q,\omega)\Pi_{0} (q,\omega)$ the
effective dynamical dielectric function. $V_{c}$, $V_{ph}$ and
$\Pi_{0}(q,\omega)$ are direct Coulomb interaction, the longitudinal
optical phonon mediated electron-electron interaction, and
irreducible polarizability,  respectively. The self-energy can be
separated into the frequency-independent exchange and correlation
parts
\begin{equation}
\label{math:2.2} \Sigma_{i}(k,\omega) =
\Sigma_{i}^{Ex}(k)+\Sigma_{i}^{Cor}(k,\omega),\quad\quad i=e,h
\end{equation}
where,
\begin{equation}
\label{math:2.2} \Sigma^{Ex}_{i}(k)=-\int^{\infty}_{-\infty}\frac{d
q}{2\pi}n_{F}(k+q)V_{s}(q).
\end{equation}
Here, $ n_{F}(k+q)=\theta(k_{F}-|k+q|)$ is the Fermi function at
$T=0$ and $\theta$ is Heaviside step function. In GW approximation,
the $\Sigma^{Cor}_{i}(k, \omega)$ can be written in the following
form,
\begin{equation}
\label{math:2.2} \Sigma^{Cor}_{i}(k, \omega)=\Sigma^{line}_{i}(k,
\omega)+\Sigma^{pole}_{i}(k, \omega),
\end{equation}
where
\begin{equation}
\label{math:2.2}
\Sigma^{line}_{i}(k,\omega)=-\int^{\infty}_{-\infty}
\frac{dq}{2\pi}V_{s}(q) \int^{\infty}_{-\infty}
\frac{d\omega'}{2\pi}\Bigg(\frac{1}{(\xi_{k+q}-\omega)-i\omega'}\Bigg)
\Bigg[\frac{1}{\epsilon(q,i\omega')}-1\Bigg],
\end{equation}
and
\begin{equation}
\label{math:2.12}
\Sigma^{pole}_{i}(k,\omega)=\int^{\infty}_{-\infty}\frac{d
q}{2\pi}\Bigg[\theta(\omega-\xi_{k+q})-\theta(-\xi_{k+q})\Bigg]
V_{s}(q)\Bigg[\frac{1}{\epsilon(q,\xi_{k+q}-\omega)}-1\Bigg]
\end{equation}
where $\xi(k)$ is defined as
\begin{equation}
\label{math:2.13} \xi(k)=\frac{\hbar^{2}k^{2}}{2m^{*}}-\mu,
\end{equation}
and $m^{*}$ is effective
electron mass[11].\\
In general the effective quasi-one dimensional Coulomb interaction
in the framework of dynamical random phase approximation is given by
\begin{eqnarray}
\label{math:2.5} V_{s}(k)&=&\int_{-\infty}^{+\infty}dy
\int_{-\infty}^{+\infty}dy'\int_{-\infty}^{+\infty}dx e^{ikx}
\frac{e^{2}}{\epsilon_{0} [(x-x')^{2}+(y-y')^{2} ]^{\frac{1}{2}}}
|\varphi(y)|^{2}|\varphi(y')|^{2} \nonumber
\\
~&~&~\nonumber \\
&=&\int_{-\infty}^{+\infty}dy \int_{-\infty}^{+\infty}dy'\nu(k,
y-y')|\varphi(y)|^{2}|\varphi(y')|^{2}.
\end{eqnarray}
In this equation $\varphi(y)$ is transverse wave function and
$\nu(k, y-y') = \frac{2e^{2}}{\epsilon_{0}}K_{0}[|k(y-y')|]$ where
$K_{0}(x)$ is zeroth order modified Bessel function of second kind.
As we will show, in our situation the problem is effectively
two dimensional(quantum well wire structure).\\

\section{BGR in V-Shaped Quantum Wires}
Using formalism of the last section, here we calculate density of
states(profile of charges distribution) and band gap
re-normalization in the geometry of V-grooved quantum quantum wire.
There are many possibilities to define mathematically the geometry
of V-grooved confinement potential[3,12]. These possibilities have
originated from different fabrication and growing scheme. A possible
and in some sense, technically suitable form of V-shaped confinement
potential can be written as,
\begin{equation}
\label{math:1.1} V(x,y)=\left\{\begin{array}{ll} 0&{\rm
if}\left\{\begin{array}{ll}-w_{x}\leq x\leq w_{x}\\\alpha x\leq
y\leq\alpha
x+w_{y}\\-\alpha x\leq y\leq-\alpha x+w_{y}\end{array}\right\}\\
\newline
\infty&{\rm elsewhere}\end{array}\right\}.
\end{equation}
For simplicity in numerical calculations, we define re-scaled
quantities $\tilde{x}=\frac{x}{w_{x}}$, $\tilde{y}=\frac{y}{w_{y}}$
and $\tilde{\alpha}=\frac{w_{x} \alpha}{w_{y}}$ for $x$, $y$
 and $\alpha$ respectively. So the re-scaled confinement potential becomes,
\begin{equation}
\label{math:1.2}\tilde{V}(x,y)=\left\{\begin{array}{ll} 0&{\rm
if}\left\{\begin{array}{ll}-1\leq \tilde{x}\leq 1\\\tilde{\alpha}
\tilde{x}\leq \tilde{y}\leq\tilde{\alpha}
\tilde{x}+1\\-\tilde{\alpha} \tilde{x}\leq \tilde{y}\leq-\tilde{\alpha} \tilde{x}+1\end{array}\right\}\\
\newline
\infty&{\rm elsewhere.}\end{array}\right\}
\end{equation}
With this form of re-scaled confinement potential, one can compute
band gap narrowing in V-grooved quantum wire. To do this end, one
should calculate effective two-dimensional transverse wave function.
According to above geometry, z-direction is without confinement, so
carriers move in this direction freely. Now the Schr\"{o}dinger
equation for carrier in two direction of confinement is, \be
\label{math:2.8} (-\frac{\hbar^2}{2m})(\frac{\partial ^2}{\partial
x^2}+\frac{\partial^2}{\partial
y^2})\phi(x,y)+V(x,y)\phi(x,y)=E\phi(x,y). \ee This can be written,
using re-scaled quantities, as \be \label{math:2.8}
(-\frac{\hbar^2}{2m})(\frac{1}{w_{x}^2}\frac{\partial^2}{\partial\tilde{x}^2}+\frac{1}{w_{y}^2}
\frac{\partial^2}{\partial\tilde{y}^2})\phi(x,y)+V(x,y)\phi(x,y)=E\phi(x,y).
\ee
Defining, $\tilde{E}=\frac{2mEw_{y}^{2}}{\hbar^2}$,
$\tilde{V}=\frac{2mVw_{y}^{2}}{\hbar^2}$, and
$c=\frac{w_{y}}{w_{x}}$, we find finally \be \label{math:2.9}
(c^2\frac{\partial^2}{\partial\tilde{x}^2}+\frac{\partial^2}{\partial\tilde{y}^{2}})\phi(x,y)-
\frac{2mV(x,y)w_{y}^{2}}{\hbar^2}\phi(x,y)=\frac{-2mEw_{y}^{2}}{\hbar^2}\phi(x,y).
\ee This equation with potential as (10), can be solved
analytically. The screened coulomb potential now is given by, \be
\label{math:2.10} V_{s}(k)=\frac{2e^2}{\varepsilon_{0}}\int dx
dy\int dx'dy' K_{0}|k \Delta r| |\varphi(x,y)|^2
 |\varphi(x',y')|^2,
 \ee
 or
 \be
\label{math:2.10} V_{s}(k)=\lambda\frac{2e^2}{\varepsilon_{0}}
w_{x}^2w_{y}^2\int dx' dy'\int d\tilde{x'}d\tilde{y'} K_{0}|k \Delta
r| |\varphi(x,y)|^2
 |\varphi(x',y')|^2,
 \ee
where $\lambda$ is re-scaling factor equal to $10^{-18}/m^2$ and $k
\Delta r =k\sqrt{(x-x_{0})^2+(y-y_{0})^2}$. Since we want to solve
the integral of equation (16), and this integral can not be solved
analytically, we need to compute it numerically and therefore the
ground state energy and ground state wave function of
Schr\"{o}dinger equation should be calculated numerically.
Therefore, we first provide a suitable numerical scheme for
calculations.\\
Basically, we have to solve 2D-Schr\"{o}dinger equation (13) for
potential as given in equation (10), which is not, theoretically,
solvable yet. Therefore, we are going to calculate the ground state
wave function numerically. We use the usual finite difference
algorithm to solve this eigenvalue problem[20]. First, we do define
the re-scaled values for $x,y,E$ and $V$ respectively as
$\frac{x}{w_x},\frac{y}{w_y},\frac{2mEw_y^2}{\hbar^2}$ and
$\frac{2mVw_{y}^2}{\hbar^2}$. Then, we do discritize the $x,y$
axes's to discritized space $dx=0.1$ and $dy=0.1$, therefore the
equation (13) will now be changed to form of the matrix equation as:
\begin{equation}
\label{math:2.10} {\cal H} \phi = E \phi,
\end{equation}
where  $\cal H$, and $\phi$ are the Hamiltonian matrix and state
wave function array respectively and are defined as: \be
\label{math:2.11}
 \phi =
 \left(\begin{array}{c}
\left(\vdots\right) \\
\left(\begin{array}{c}
\vdots \\
\phi(x_i,y_j) \\
\phi(x_{i+1},y_j) \\
\vdots
\end{array}
\right ) \\
\left(\begin{array}{c}
\vdots \\
\phi(x_i,y_{j+1}) \\
\phi(x_{i+1},y_{j+1}) \\
\vdots
\end{array}
\right ) \\
\left(\vdots\right)
\end{array}
\right ), \ee and
\be {\cal H}=\left(\begin{array}{cccc}
  \left(\ddots \right) & \left(\begin{array}{ccc}
  \ddots & 0 & 0   \\
  0 & -\frac{1}{dy^2}& 0  \\
  0 & 0 & \ddots
  \end{array}\right) & 0 & 0  \\
  \left(\ddots \right) &
  \left(\begin{array}{cccc}
  \ddots & \ddots & 0 & 0  \\
  \ddots &2(\frac{c^2}{dx^2}+\frac{1}{dy^2})+V(x_i,y_j) & \frac{-c^2}{dx^2} & 0 \\
  0 & \frac{-c^2}{dx^2} & 2(\frac{c^2}{dx^2}+\frac{1}{dy^2})+V(x_{i+1},y_j) & \ddots \\
  0 & 0 & \ddots & \ddots
  \end{array}\right)
  & \left(\ddots\right)
  & 0   \\
  0 &
\left(\begin{array}{ccc}
  \ddots & 0 & 0   \\
  0 & -\frac{1}{dy^2}& 0  \\
  0 & 0 & \ddots
  \end{array}\right)
    & \left(\ddots\right)& \left(\ddots\right) \\
  0 & 0 & \left(\ddots \right)& \left(\ddots\right)
\end{array} \right )
\ee The error of computation of $\phi(x,y)$ is in the order of
$O(dx^2)$. We diagonalize the matrix and calculate the ground state
wave function numerically. The screened potential is calculated by
ground state wave function as a function of k. Figure 1 shows the
calculated screened potential versus the wave vector for different
$\alpha$. In this figure, $V_s(k)$ is normalized by
$\frac{2e^2}{\epsilon_0}$ and the $k$ is normalized to $kw_y$. Now,
we are able to calculate the self energies of electrons and holes.
For this goal, first we define the re-scaled $\beta_e$ and $\mu_e^0$
respectively as $\frac{\beta \hbar^2}{2m_e^* w_y^2}=\beta
\frac{574.5}{w_y^2}$ and $\mu^0_e \beta$, where $w_y$ is width of
quantum well in $y$ direction in Nanometer. For holes, we also
define the re-scaled $\beta_h$ and $\mu^0_h$ as $\beta_h =\beta_e
\big(\frac{m^*_e}{m^*_h} \big)$ and $\mu_h^0=\mu_e^0
\big(\frac{m^*_e}{m^*_h} \big)$ respectively. In all computations in
this paper, we have assumed that the ratio $\frac{m_h^*}{m_e^*}$ is
equal to $0.3$ and $m_e^*\simeq 0.067 m_e$. By calculating the self
energies of holes and electrons, we compute the band gap
re-normalized energy for V-grooved confinement potential in
dynamical random phase(GW) approximation. We consider the case of
$T=0$. The result is shown in figure 2 which shows the values of
bandgap renormalization in terms of the width of the V-shaped
confinement potential. In all figures, we have assumed that chemical
potentials for electron is $1 meV$ and $ E_g$ is normalized by
$2e^2/\epsilon_0$.
\section{BGR in T-Shaped Quantum Wires}
Now we want to do the same calculations for geometry of T-shaped
quantum wires. It is evident that one can consider the external
T-shaped confinement potential as follows, \be \label{math:2.7}
V(x,y)=\left\{
\begin{array}{cc}
0 & \hspace{2 cm}-\infty< x < \infty, \quad \frac{-w_{y}}{2}\leq y \leq \frac{w_{y}}{2}\\

0  &\hspace{2 cm}\frac{-w_{y}}{2}< y < \infty, \quad \frac{-w_{x}}{2}\leq x \leq \frac{w_{x}}{2}\\

\infty & \hspace{3cm}elsewhere.
\end{array}
\right \} \ee Note that we want to simulate actual situation with
mathematical functions. The actual problem is a quantum wire which
is a quasi-one dimensional structure but its external confinement
potential should be effectively two dimensional to provide two
direction of confinement. Therefore, this form of confinement
potential is considered effectively two dimensional. To proceed the
same algorithm as preceding section, we first re-scale some
quantities. First of all the confinement potential can be written
as, \be \label{math:2.9} \widetilde{V}=\left\{
\begin{array}{cc}
0 &\hspace{2 cm}-\infty< x < \infty, \quad \frac{-1}{2}\leq y \leq \frac{1}{2}\\
0 &\hspace{2 cm}\frac{-1}{2}< y < \infty, \quad \frac{-1}{2}\leq x \leq \frac{1}{2}\\
\infty &\hspace{3cm} elsewhere.
\end{array}
\right\} \ee The re-scaled Schr\"{o}dinger equation becomes
\begin{equation}
\label{math:2.10} -(\frac{\partial^2}{\partial
x^2}+\alpha^2\frac{\partial^2}{\partial y^2})\phi(x,y)+\frac{2m
\widetilde{V}(x,y)}{\hbar^2w_y^2}\phi(x,y)=\frac{2mE}{\hbar^2w_y^2}\phi(x,y),
\end{equation}
where now $\alpha$ is ratio of well width potential in $x$ direction
to the $y$ direction, i.e. $\alpha=\frac{w_x}{w_y}$. The same
numerical scheme has been used to calculate various quantities. By
diagonalization of the Hamiltonian matrix, we calculate the ground
state wave function numerically. Using ground state wave function,
one can calculate the concentration of carriers in quantum wire
structure and therefore density of probability for these carriers.
This procedure is as previous section. One can show that by
decreasing $\alpha$, the wave function of carriers will be localized
in $x$ directions while by increasing $\alpha$ the wave function
will be more localized in $y$ directions of the well relative to $x$
direction. From physical grounds, this is reasonable to say that
probability density and charge
distribution are function of external confinement potential.\\
Now we should calculate the screened Coulomb potential. In our
situation the screened Coulomb potential should be considered more
carefully, since now the confinement potential of quantum well wire
is effectively two dimensional. It is important to note that this
means that quantum mechanical problem under consideration is
effectively two-dimensional. For these reasons, the screened Coulomb
potential can be written as \be \label{math:2.6} V_{s}(k)
=\int_{-\infty}^{+\infty}dxdy\int_{-\infty}^{+\infty}dx'dy'\nu(k,\rho-\rho')
|\varphi(x,y)|^{2}|\varphi(x',y')|^{2} \ee where
$\rho=|x\hat{i}+y\hat{j}|$ is the radius polar coordinate in plane
and $\varphi(x,y)$ is transverse wave function and $\nu(k,
\rho-\rho') = \frac{2e^{2}}{\epsilon_{0}}K_{0}[|k(\rho-\rho')|]$
where $K_{0}(x)$ is zeroth order modified Bessel function of second
kind. With this point in mind the screened Coulomb potential (16) is
calculated by two dimensional ground state wave function. We have
calculated the screened Coulomb potential, as a function of $k$ in
figure (3). In this figure, $V_s(k)$ is normalized by
$\frac{2e^2}{\epsilon_0}$ and the $k$ is normalized to $1/W_y$ and
the figure shows that screened potential is going to zero at lower
wave length, and it is infinite at higher wave length for different
$\alpha$ ratio. The screened potential for a constant wave vector of
carriers will increase by increasing $\alpha$. This indicates that
screening effects for
carriers are affected by the ratio of width of well potentials.\\
Now  to calculate the self energies of electrons and holes first we
define the re-scaled $\beta_e$ and $\mu_e^0$ respectively as before.
By calculating the self energies of holes and electrons, we compute
the band gap energy renormalization  for different temperatures and
width of the quantum wells for T-shaped potential. Figure (4) shows
the relative difference of band gap energy renormalization for
T-shaped potential versus the width of quantum well in $y$
directions. Increasing temperature leads to more relative
renormalization of gap energy and this is in agreement with the
experimental results[1,13]. Also it is evident that in our situation
the relative width of the well, i.e. $\alpha$ has significant effect
on the value of band gap renormalization. Also our analysis shows a
temperature dependence of gap energy renormalization. Generally,
under high optical excitation, the band gap for $2D$ and bulk
systems is found to decrease with plasma density due to
exchange-correlation effects. The observed band gaps are typically
normalized by $\sim 20 meV$ within the range of plasma densities of
interest which arise from the conduction band electrons and valence
band holes. Since Coulomb screening decreases with reduction of
semiconductor dimensions, typical values of band gap renormalization
in the case of bulk semiconductor is more than gap renormalization
in quantum well structures. In the same manner typical band gap
renormalization for quantum wires is less than that of quantum
wells. Our numerical results, when are compared with experimental
data[13], show the good agreement with experiments at least in
general behavior.

\section{Results and Discussions}
In this paper, a numerical approach has been proposed to calculate
band gap renormalization in a V-shaped and T-shaped quantum wires.
Quantum field theoretical random phase approximation within the
leading-order screening has been used to calculate screened Coulomb
potential in terms of carrier density and width of the confinement
potentials. The essential question is about the form of confinement
potentials. As we have argued the quantum mechanical problem under
consideration is effectively two dimensional. In other words since
we have two dimension of confinement, simulation of external
confinement with mathematical functions, leads to an effectively two
dimensional confinement potential which requires solution of two
dimensional Schr\"{o}dinger equation. Based on this viewpoint we
have shown that using two dimensional Cartesian and polar
coordinates for V-shaped and T-shaped quantum wire respectively, one
can use the formalism of Hu and Das Sarma with re-definition of some
quantities. Our numerical calculations show that the distribution of
probability for carriers concentration is a function of the ratio of
the well width in two dimensions. This is natural since probability
distributions is a function of confinement potential. In other
words, the geometric shape of external confinement itself restricts
the shape of wave functions. As the figures show, by decreasing
$\alpha$, the wave function of carriers will be localized in $x$
directions while by increasing $\alpha$ the wave function will be
more localized in $y$ directions of the well relative to $x$
direction. It is important to note that by re-scaling procedure
which we have considered, we have fixed the geometry of the wires
but now Schr\"{o}dinger equation becomes re-scaled via the presence
of ratio $\alpha$. Actually one should consider the possibility for
changing the geometry also. But our investigation show that the main
physical results do not change considerably. Calculation of screened
Coulomb interaction shows that the screened potential is going to
zero at lower wave length, and it is infinite at higher wave length
for different $\alpha$ ratio. The screened potential for a constant
wave vector of carriers will increase by increasing $\alpha$. This
indicates that screening effects for carriers are affected by the
ratio of width of well potentials. Also calculation of band gap
renormalization shows that increasing temperature leads to more
relative renormalization of gap energy and this is in agreement with
the experimental results. Also it is evident that in our situation
the relative width of the well, i.e. $\alpha$ has significant effect
on the value of band gap renormalization. This is not surprising
since the screening effect itself is dependent on the shape of
confinement potential. Since Coulomb screening decreases with
reduction of semiconductor dimensions, typical values of band gap
renormalization in the case of bulk semiconductor is more than
quantum well structures. In the same manner typical band gap
renormalization for quantum wires is less than that of quantum
wells. The relative values of band gap renormalization in various
fabrication pattern(i.e. V-shaped, T-shaped and rectangular) quantum
wire structure is dependent to the shape and depth of the potential
well. Our numerical results, when are compared with experimental
data[13], show the good agreement with experiments at least in
general behavior.

\input{epsf}
\begin{figure}[ht]
\begin{center}
\includegraphics{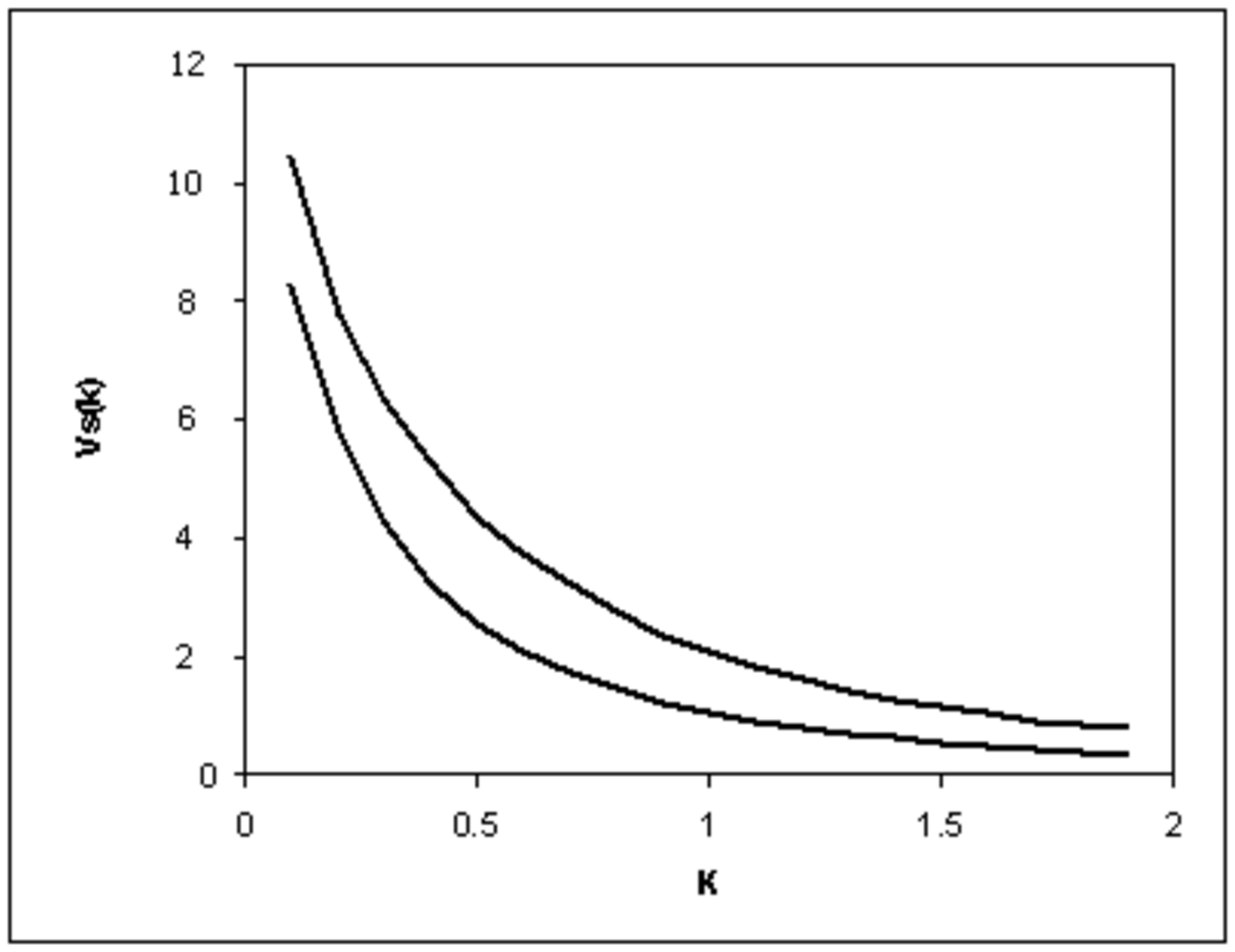}

\end{center}
\vspace{10 cm}
 \caption{\small {The calculated screened potential versus the wave vector for different $\alpha$,
 where we have set $\alpha= 0.5$ for lower curve and $\alpha=1.0$ for upper one.
 The screened potential is normalized by
$\frac{2e^2}{\epsilon_0}$ and the $k$ is normalized to $kw_y$. }}
 \label{Fig:1}
\end{figure}

\begin{figure}[ht]
\begin{center}
\includegraphics{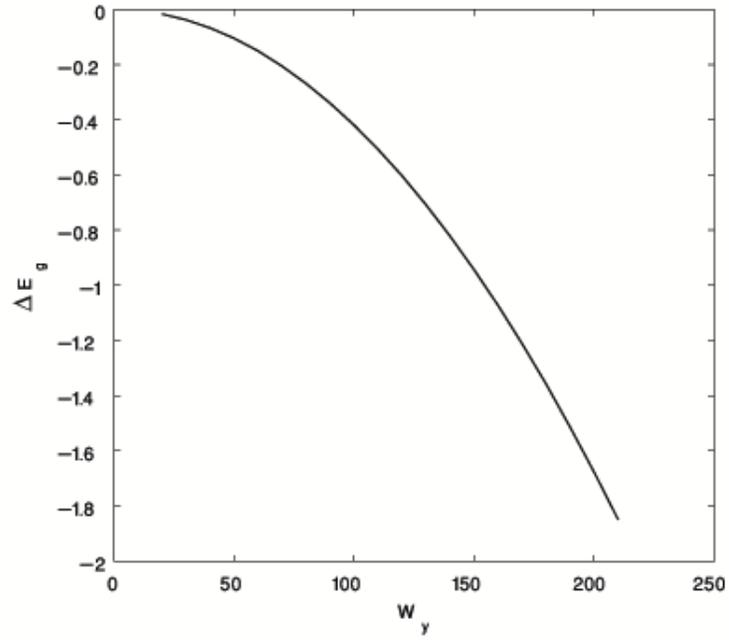}
\end{center}
\vspace{10 cm} \caption{\small { Renormalization of gap energy
calculated numerically for V-shaped potential versus the width of
quantum well wire $w_y$. In all figures, we have assumed that
chemical potentials for electron is $1 meV$. The $\Delta E_g$ is
normalized by $2e^2/\epsilon_0$}.} \label{Fig:3}
\end{figure}
\input{epsf}

\begin{figure}[ht]
\begin{center}
\includegraphics{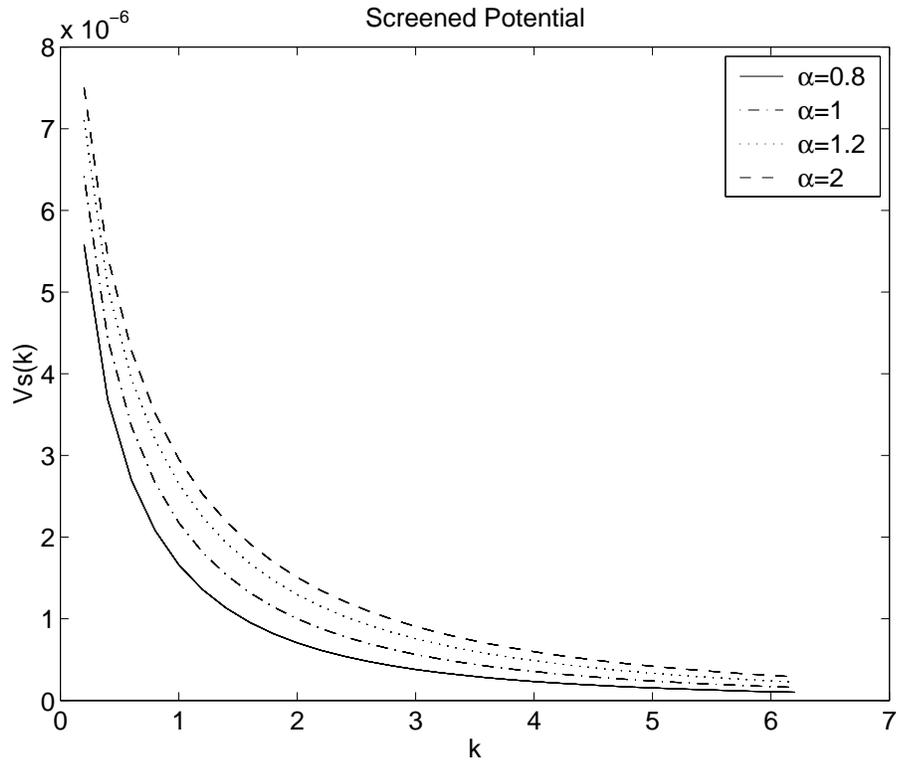}

\end{center}
\vspace{10 cm}
 \caption{\small {The calculated screened Coulomb potential versus the wave vector.
 The screened potential is normalized by
$\frac{2e^2}{\epsilon_0}$ and the $k$ is normalized to $1/w_y$. }}
\end{figure}

\begin{figure}[ht]
\begin{center}
\includegraphics{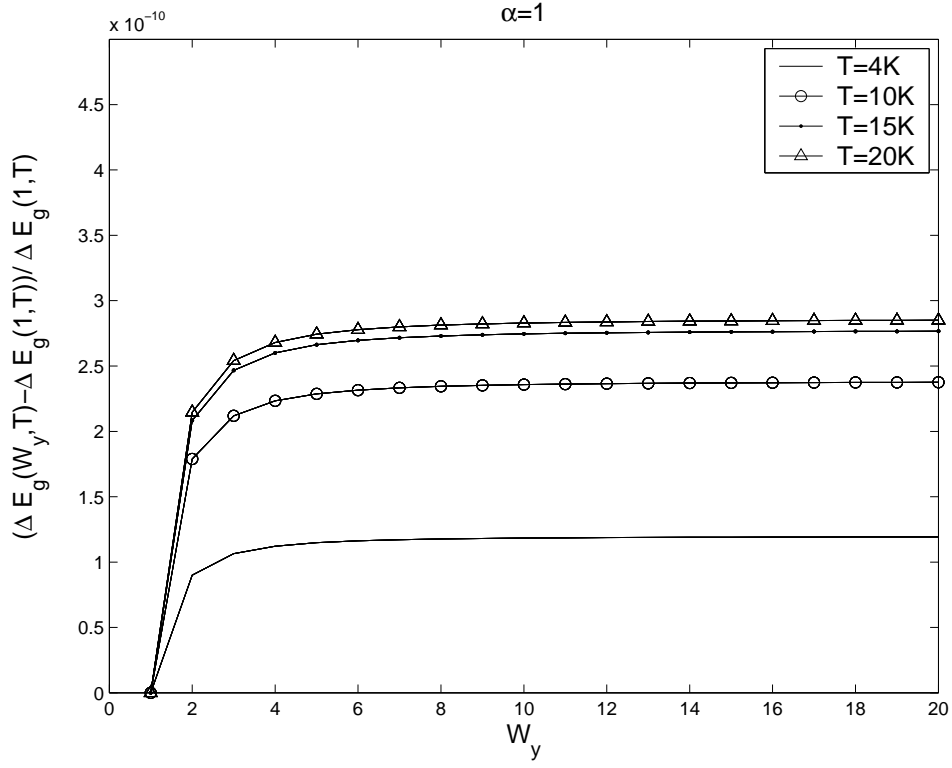}
\end{center}
\vspace{10 cm} \caption{\small { This figure shows the difference of
band gap renormalized energy for T-shaped confinement potential
versus the width of quantum well in $y$ directions for different
temperatures and different ratio of the well width in two
dimensions, $\alpha$. In all figures, we have assumed that the
chemical potentials for electron is  $1 meV$. The $\Delta E_g$ is
normalized by $2e^2/\epsilon_0$}.}
\end{figure}

\end{document}